# CANVAS: Captioning Art with Narrative Visual-Audio AI Systems

**Vignesh Nagarajan**
Texas A&M University
Department of Computer Science & Engineering
College Station, TX, USA
vigneshn26@tamu.edu

**Abstract** For a person who is blind or has low vision (BLV), a painting is usually represented by a few words of alt-text, if anything at all. That text rarely conveys the colour, texture, mood or arrangement that a sighted visitor takes in, and rich audio descriptions are too slow and costly to produce by hand for most collections. This study asks how far off-the-shelf generative AI can close that gap. We built CANVAS, a no-code Zapier workflow that sends an artwork image to a multimodal large language model, checks the returned description, narrates it with a fixed ElevenLabs voice and logs every run. With prompt, pipeline and voice held constant, Gemini 2.5 Flash, GPT-5 Mini and Claude 3.5 Haiku each described the same 50 artworks, and we compared the 150 narrated descriptions artwork by artwork. Gemini and GPT-5 Mini wrote text of similar Flesch Reading Ease (49.95 and 49.47); Claude's was harder to read (37.35, $p < 0.001$), largely because it used longer words. Lexical diversity did not differ. The findings show both promise and a clear design requirement. All 150 narrated descriptions cost US$5.10, so a first draft of audio description for an entire collection is affordable even for small museums and educators. Yet while every description conveyed colour, only 16–18% used explicit layout terms such as left, right, foreground or background, and at most 10% signalled uncertainty. Spatial structure must therefore be requested explicitly and checked by people; accuracy and usefulness to BLV listeners remain to be tested.

## 1 Introduction

Most visual art is effectively inaccessible to people who are blind or have low vision. Museums and online collections usually rely on short alt-text or on audio guides written for sighted visitors, and neither conveys much about how a work looks: its composition, materials, light or mood. Doore et al. [1] found that even carefully written curator descriptions did not explicitly convey the spatial information a BLV visitor needs to build a mental model of a scene. Interviews and a large survey of blind arts patrons describe both the value they place on verbal description and the limits of what is currently offered [2]. Research on museum audio description argues that good description goes beyond listing objects to convey style, emotion and context, sometimes through sensory metaphor [3, 4].

Automatic image description has improved quickly. Encoder–decoder captioners with visual attention [5, 6] and paragraph-level generators [7] have given way to multimodal LLMs that write fluent, multi-sentence descriptions on demand. These systems were not designed around BLV needs, however. Models trained mostly on everyday photographs perform worse on artworks [8], and generic captions emphasise object labels over layout, technique and mood. Studies with BLV users add two cautions. First, what a person wants from a description depends on context, so one-size-fits-all descriptions fall short [9]. Second, blind users tend to trust fluent captions and fill in missing detail, which makes confident errors costly; MacLeod et al. [10] recommend wording that signals uncertainty. Recent work with large vision–language models finds that BLV users' ratings of model descriptions vary widely in whether they give enough detail and whether they are concise, and that no single model is consistently preferred [11]. Common automatic metrics also correlate poorly with what BLV users find helpful [11, 12].

This article asks a practical question: if a museum or educator assembled an end-to-end pipeline from off-the-shelf services, what would the resulting descriptions look like, and what would they cost? We built CANVAS (Captioning Art with Narrative Visual-Audio AI Systems), a no-code workflow that turns an uploaded artwork image into a short narrated description, and used it to compare three contemporary multimodal LLMs on the same 50 artworks. The study addresses three research questions:

- **RQ1.** Under an identical prompt and pipeline, how do the three models differ in the readability and sentence structure of their descriptions?
- **RQ2.** How often do the descriptions contain the content BLV-oriented guidance emphasises (colour, texture, mood, medium and especially spatial layout), signal uncertainty, or rely on recognising the specific work?

- **RQ3.** What time and cost does the pipeline incur, separating recorded values from estimates?

The study makes four contributions. (1) It documents a reproducible low-code pipeline for narrated artwork descriptions and releases all 150 descriptions, audio links, scores and analysis code [13]. (2) It reports a paired, within-artwork comparison of three models, showing that their readability differences arise from different mechanisms and that their lexical richness does not differ. (3) Its content analysis shows a consistent gap in spatial information and uncertainty signalling across all three models, which translates into concrete prompt and review requirements. (4) It separates recorded cost and time from estimates of manual production. The study does not evaluate accuracy or usefulness with BLV listeners; Sect. 8 discusses what that implies.

## 2 Related work

### *2.1 Art access for BLV audiences*

Museum audio description has its own literature, distinct from general alt-text. Hutchinson and Eardley [4] examine the problem of textual fidelity in museum audio description, that is, how closely and in what form a description should follow the work it describes. Neves [3] argues for multi-sensory description that uses subjective interpretation and analogies to touch and sound so that BLV visitors can engage with art aesthetically rather than only factually. Doore et al. [1] analysed descriptions of artworks and highlighted the absence of explicit spatial information as a recurring weakness. Li et al. [2] combined interviews with a survey of 220 blind arts patrons and derived eight design considerations for technology that supports access to visual art. Together these studies suggest that a useful description of an artwork should cover what is depicted, where things are, what the work is made of and how it feels, while being clear about what is observation and what is interpretation.

### *2.2 Image description for accessibility and its evaluation*

Image descriptions written for BLV users differ from generic captions. Stangl et al. [9] showed that the information people want depends on the scenario and the source of the image. Gurari et al. [14] built a captioning dataset from photographs taken by blind people and showed that models trained on conventional datasets transfer poorly to it. In this journal, Leotta et al. [15] evaluated automatic captioning for web accessibility and found that automatically generated descriptions are often inadequate as alternative text. MacLeod et al. [10] documented how blind users over-trust captions phrased with certainty. With multimodal LLMs, BLV people have begun using scene-description apps for many everyday tasks [16], and An et al. [11] found that BLV participants' preferences among model-generated descriptions are only weakly captured by automatic metrics. Kreiss et al. [12] likewise showed that referenceless metrics ignore context in ways that matter for accessibility. Accessibility standards bear on the language of descriptions as well as their content: at its highest conformance level, WCAG 2.1 asks for a simpler supplementary version whenever text requires reading ability above lower secondary education [17]. These findings motivate evaluating descriptions against the content BLV users ask for rather than against surface similarity to reference captions.

### *2.3 Captioning models, art and hallucination*

Surveys of image description trace the field from template and retrieval methods to neural encoder–decoder models [18]. Attention over image regions [5, 6] and hierarchical paragraph generation [7] improved coverage of salient content. Art poses its own problems: domain shift from photographs [8], and the affective and interpretive character of what people say about art, which ArtEmis captured at scale [19]. A further risk is hallucination, meaning content that is fluent but not supported by the image. It was first quantified for object mentions in captioning [20] and has been studied extensively for large vision–language models [21, 22]. Well-known artworks add a further route to unsupported content: a model may recognise a famous work and describe it from memory instead of from the pixels it was given.

Prior work therefore either studies what BLV audiences need or develops and benchmarks captioning models. Few studies examine what an off-the-shelf, end-to-end pipeline using current commercial models produces for artworks, measured with a paired design and against BLV-relevant content. This article fills that gap descriptively; it complements rather than replaces user studies.

## 3 The CANVAS pipeline

CANVAS is implemented as a Zapier workflow (a "Zap"), shown in Fig. 1. We built three Zaps that are identical except for the model selected in the language-model step, so that the model is the only experimental variable. We used a no-code orchestrator because it makes every step and branch explicit, lets each block (trigger, generation, validation, synthesis, logging) be replaced independently, and can be rebuilt by museum or education staff without software-engineering support.

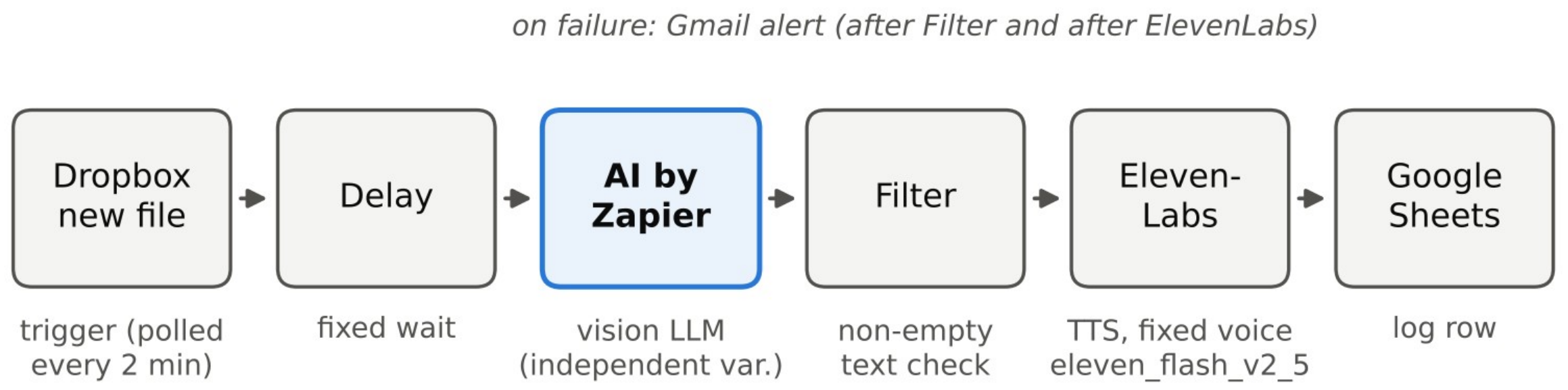

**Fig. 1** The CANVAS workflow. Each of the three Zaps is identical except for the model chosen in the AI by Zapier step. Gmail alerts are sent if the Filter step finds no description or if text-to-speech fails.

**Trigger and delay.** A Dropbox "New File in Folder" trigger starts the Zap when an image is added to a designated folder. Zapier checks the folder by polling, at a two-minute interval for this trigger, so the time from upload to processing depends on when the next poll occurs. A fixed Delay step follows so that the file is fully synchronised before it is read; its duration is identical in all three Zaps.

**Description.** The "AI by Zapier: Analyze and Return Data" step receives the Dropbox file as an image input and sends it, with the prompt below, to the selected model: Gemini 2.5 Flash, GPT-5 Mini or Claude 3.5 Haiku, as labelled in Zapier. The prompt was identical in all three Zaps:

```
You are a sensory-focused visual describer. Look at {{1. Dropbox File}} and write
a short, vivid, 2-3 sentence description that includes colors, textures, mood,
and atmosphere. Avoid technical jargon. No more than 90 words. Don't be overly
descriptive, and be sure to clearly describe what the art is about.
```

The prompt asks for colours, textures, mood and atmosphere and caps length at 90 words. It does not ask for spatial layout. Decoding parameters were left at the AI by Zapier defaults, which were the same for every run; the action does not expose provider-side settings or model snapshot identifiers, so the model names above are the platform's labels.

**Validation and narration.** A Filter step passes the run on only if a non-empty text description was returned; otherwise a Gmail alert records the file name and response. This check catches failures but not errors in content: a fluent but wrong description passes. Valid descriptions are sent to ElevenLabs text-to-speech with a single voice and model (eleven_flash_v2_5) for every run, so that voice and synthesis settings cannot differ between conditions. A second Gmail alert fires if synthesis fails.

**Logging.** Each successful run appends a row to a Google Sheet with the artwork name, image URL, description, audio file URL, ElevenLabs audio ID, ElevenLabs model and voice IDs. Readability and the other measures in Sect. 5 were computed afterwards from these logged descriptions; nothing was edited before scoring.

## 4 Dataset

The input set contains 50 well-known artworks in five curatorial groups of ten: (1) Renaissance and Baroque, (2) Impressionism and its successors, (3) Modern and abstract, (4) Photography, sculpture, installation and mixed media, and (5) Narrative scenes and landscapes. Appendix A lists every work. The groups are curatorial rather than art-historical; for example, group 4 also contains three paintings (Manet, Lichtenstein, Kahlo) and a woodblock print (Hokusai). Images were downloaded from Wikimedia Commons and museum open-access collections and used without cropping, resizing or colour adjustment.

Three properties of the dataset matter for interpretation. First, the works are among the most reproduced in the world, and each file name contains the work's title (e.g. Las-Meninas.jpg). A model may therefore recognise a work, from the image or the file name, and draw on prior knowledge; Sect. 6.4 measures how often descriptions show this. Second, inputs were identical across models for 49 of 50 works; for *Las Meninas*, Gemini 2.5 Flash received a JPEG and the other two models a WebP file. Excluding this work does not change the readability comparisons (e.g. the Gemini–Claude difference in Flesch Reading Ease becomes 12.23 instead of 12.60 points). Third, 36 works are in the public domain in most jurisdictions, but 14 are by artists who died in 1954 or later or are still living (marked in Appendix A) and are likely still protected by copyright in at least some jurisdictions. These images were used only as model inputs for research and are not reproduced in this article.

## 5 Measures and analysis

### *5.1 Readability and structure*

Readability was measured with the Flesch Reading Ease score (FRE) [23], FRE = 206.835 − 1.015 × (words / sentences) − 84.6 × (syllables / words), where higher values indicate easier text. Scores were computed with a short Python script using a heuristic syllable counter and are analysed as originally recorded. Recomputing them with the archived script reproduces 132 of the 150 values exactly; the other 18 differ by a median of 0.84 points (maximum 6.72) because of how apostrophes are tokenised. As a sensitivity check, all FRE analyses were repeated with the recomputed values and with an independent implementation (textstat 0.7.3). We also report the two components of FRE, average sentence length (words per sentence) and average word length (syllables per word), together with word and sentence counts. FRE measures reading difficulty, not listening difficulty, and is coarse for texts of two or three sentences; we therefore use it to compare models, not to assign grade levels.

### *5.2 Lexical richness*

Lexical diversity was measured with the moving-average type–token ratio over a 25-token window (MATTR) [24] and with MTLD [25]; both are less sensitive to text length than a raw type–token ratio. Adjective density is the proportion of alphabetic tokens tagged as adjectives by spaCy 3.8 with the en_core_web_sm 3.8.0 model [26].

### *5.3 Content coverage*

To relate the descriptions to the content BLV-oriented guidance emphasises, each description was coded automatically for the presence of at least one term from each of six dictionaries: colour; texture or material; mood or atmosphere; medium or form (e.g. painting, photograph, triptych); spatial or relational terms (e.g. behind, beneath, distant, across); and a stricter subset of explicit layout terms (left, right, centre, middle, foreground, background, top, bottom, corner). Two further dictionaries capture how claims are framed: uncertainty markers (e.g. seems, appears, may, as if) and interpretive cues (e.g. suggests, evokes, hints). Appendix B lists every term. These are lexical proxies: they show whether a kind of content is mentioned, not whether it is correct. They can also over-count. The most frequent texture term, "soft" (71 occurrences), often describes light rather than a surface, so the texture figures in particular are upper bounds.

### *5.4 Recognition of the specific work*

A description was coded as relying on recognition if it named the artist, cited the work's title, or identified depicted people, places or events that cannot be read from the image alone (e.g. "Christ and his disciples", "Socrates", "Eden"). Titles that are ordinary descriptions of what is shown (e.g. *The Scream*, *Water Lilies*, *Black Square*) were excluded, because using those words is description rather than recognition. We separately counted descriptions that call the work "iconic", "famous" or a "masterpiece".

### *5.5 Prompt compliance*

Compliance was defined as 90 words or fewer and two or three sentences, as the prompt requested.

### *5.6 Statistical analysis*

Each artwork was described by all three models, so all comparisons are within-artwork (paired). For continuous measures we used a one-way repeated-measures ANOVA with a Greenhouse–Geisser correction [27], reporting partial $\eta^2$, and a Friedman test as a non-parametric check [28]. Pairwise comparisons are paired t-tests with Holm correction [29], reported as the mean paired difference with its 95% confidence interval and Cohen's $d_z$; Wilcoxon signed-rank tests with Holm correction were run as a check and agree with the t-tests except where noted. For binary content codes we used Cochran's Q test followed by exact McNemar tests with Holm correction. For FRE we also fitted a linear mixed model with model and artwork group as fixed effects and a random intercept for each artwork, and tested a model × group interaction with a likelihood-ratio test. Analyses used Python with SciPy 1.18 and statsmodels 0.15; the code is in Online Resource 1.

### *5.7 Time and cost*

We distinguish recorded from estimated values. Recorded values are the run dates and narration timestamps (the ElevenLabs audio ID of each run embeds a millisecond timestamp) and the actual spend on each service. Upload-to-audio latency for individual images was not logged. Estimated values are the costs of producing comparable descriptions manually, which we compute from stated unit-rate assumptions.

## 6 Results

The log contains one successful run for each of the 150 artwork–model pairs. The runs took place on 5 November 2025 (Gemini 2.5 Flash, 02:35–04:52 UTC; GPT-5 Mini, 05:17–05:48 UTC) and 6 November 2025 (Claude 3.5 Haiku, 18:22–18:46 UTC). All descriptions were 90 words or fewer, and 149 of 150 had two or three sentences (one Gemini description had four).

### *6.1 Readability and structure (RQ1)*

**Table 1** Readability, structure and lexical richness by model (mean and SD; n = 50 descriptions per model)

| Measure | Gemini 2.5 Flash | GPT-5 Mini | Claude 3.5 Haiku | F(2, 98) | p | Partial $\eta^2$ |
|---|---|---|---|---|---|---|
| Flesch Reading Ease | 49.95 (11.1) | 49.47 (8.5) | 37.35 (8.4) | 47.88 | < 0.001 | 0.49 |
| Words | 55.4 (6.5) | 57.0 (8.6) | 48.5 (5.4) | 30.03 | < 0.001 | 0.38 |
| Sentences | 2.96 (0.28) | 2.04 (0.20) | 2.02 (0.14) | 273.87 | < 0.001 | 0.85 |
| Words per sentence | 18.9 (2.9) | 28.1 (4.7) | 24.1 (2.7) | 122.52 | < 0.001 | 0.71 |
| Syllables per word | 1.63 (0.13) | 1.52 (0.10) | 1.72 (0.10) | 59.95 | < 0.001 | 0.55 |
| MATTR (25-token window) | 0.912 (0.035) | 0.915 (0.029) | 0.922 (0.034) | 1.66 | 0.196 | 0.03 |
| MTLD | 105.0 (27.8) | 113.2 (27.1) | 110.0 (32.7) | 1.43 | 0.246 | 0.03 |
| Adjective density (%) | 24.7 (4.8) | 24.2 (4.7) | 23.7 (4.5) | 0.66 | 0.513 | 0.01 |

Repeated-measures ANOVA across the three models with the artwork as the repeated unit; p values are Greenhouse–Geisser corrected. Friedman tests reached the same conclusion for every measure. Pairwise comparisons are given in Table 2.

FRE differed between models (repeated-measures ANOVA, $F(2, 98) = 47.88$, $\varepsilon = 0.98$, $p < 0.001$, partial $\eta^2 = 0.49$; Friedman $\chi^2(2) = 43.36$, $p < 0.001$; Table 1 and Fig. 2). Table 2 gives every paired comparison. Gemini 2.5 Flash and GPT-5 Mini did not differ (mean paired difference 0.48 points, 95% CI −2.54 to 3.50), whereas both scored about 12 points higher than Claude 3.5 Haiku ($d_z = 1.17$ and 1.27), and did so for 42 and 46 of the 50 artworks respectively. The mixed model gave the same estimates (Claude 3.5 Haiku lower by 12.60 and 12.13 points) and showed that artworks account for 38% of the variance not explained by model and group. The model × group interaction was not significant ($\chi^2(8) = 8.04$, $p = 0.43$), so the gap between models was similar across the five groups (Table 3). The pattern was identical with the recomputed scores and with textstat (means 58.26, 59.20 and 46.68; $F(2, 98) = 41.18$, $p < 0.001$), although textstat's syllable counting raises every score by about nine points. Absolute FRE values, and the grade levels sometimes attached to them, therefore depend on implementation details, while the comparison between models does not.

**Table 2** Paired comparisons between models (n = 50 artworks; difference = first model minus second)

| Comparison | Mean difference | 95% CI | t(49) | $p_{Holm}$ | $d_z$ | Artworks higher (first / second) |
|---|---|---|---|---|---|---|
| **Flesch Reading Ease (points)** | | | | | | |
| Gemini − GPT-5 Mini | 0.48 | −2.54 to 3.50 | 0.32 | 0.752 | 0.04 | 26 / 24 |
| Gemini − Claude | 12.60 | 9.55 to 15.66 | 8.29 | < 0.001 | 1.17 | 42 / 8 |
| GPT-5 Mini − Claude | 12.13 | 9.41 to 14.84 | 8.97 | < 0.001 | 1.27 | 46 / 4 |
| **Words per sentence** | | | | | | |
| Gemini − GPT-5 Mini | −9.22 | −10.33 to −8.10 | −16.63 | < 0.001 | −2.35 | 0 / 50 |
| Gemini − Claude | −5.17 | −6.26 to −4.08 | −9.51 | < 0.001 | −1.34 | 3 / 46 |
| GPT-5 Mini − Claude | 4.05 | 2.71 to 5.39 | 6.09 | < 0.001 | 0.86 | 39 / 9 |
| **Syllables per word** | | | | | | |
| Gemini − GPT-5 Mini | 0.107 | 0.074 to 0.140 | 6.49 | < 0.001 | 0.92 | 41 / 9 |
| Gemini − Claude | −0.087 | −0.126 to −0.048 | −4.46 | < 0.001 | −0.63 | 10 / 40 |
| GPT-5 Mini − Claude | −0.194 | −0.228 to −0.159 | −11.36 | < 0.001 | −1.61 | 2 / 48 |

Paired t-tests with Holm correction within each measure. Wilcoxon signed-rank tests with Holm correction led to the same conclusion for every comparison shown. Ties are excluded from the last column.

**Table 3** Mean Flesch Reading Ease by artwork group (10 artworks per group)

| Group | Gemini 2.5 Flash | GPT-5 Mini | Claude 3.5 Haiku |
|---|---|---|---|
| 1 Renaissance and Baroque | 49.85 | 46.03 | 33.91 |
| 2 Impressionism and its successors | 46.40 | 50.09 | 38.81 |
| 3 Modern and abstract | 50.45 | 49.92 | 37.91 |
| 4 Photography, sculpture, installation, mixed media | 45.84 | 47.24 | 36.57 |
| 5 Narrative scenes and landscapes | 57.21 | 54.08 | 39.54 |

Model × group interaction in the linear mixed model: likelihood-ratio $\chi^2(8) = 8.04$, $p = 0.43$. Claude 3.5 Haiku had the lowest mean in every group.

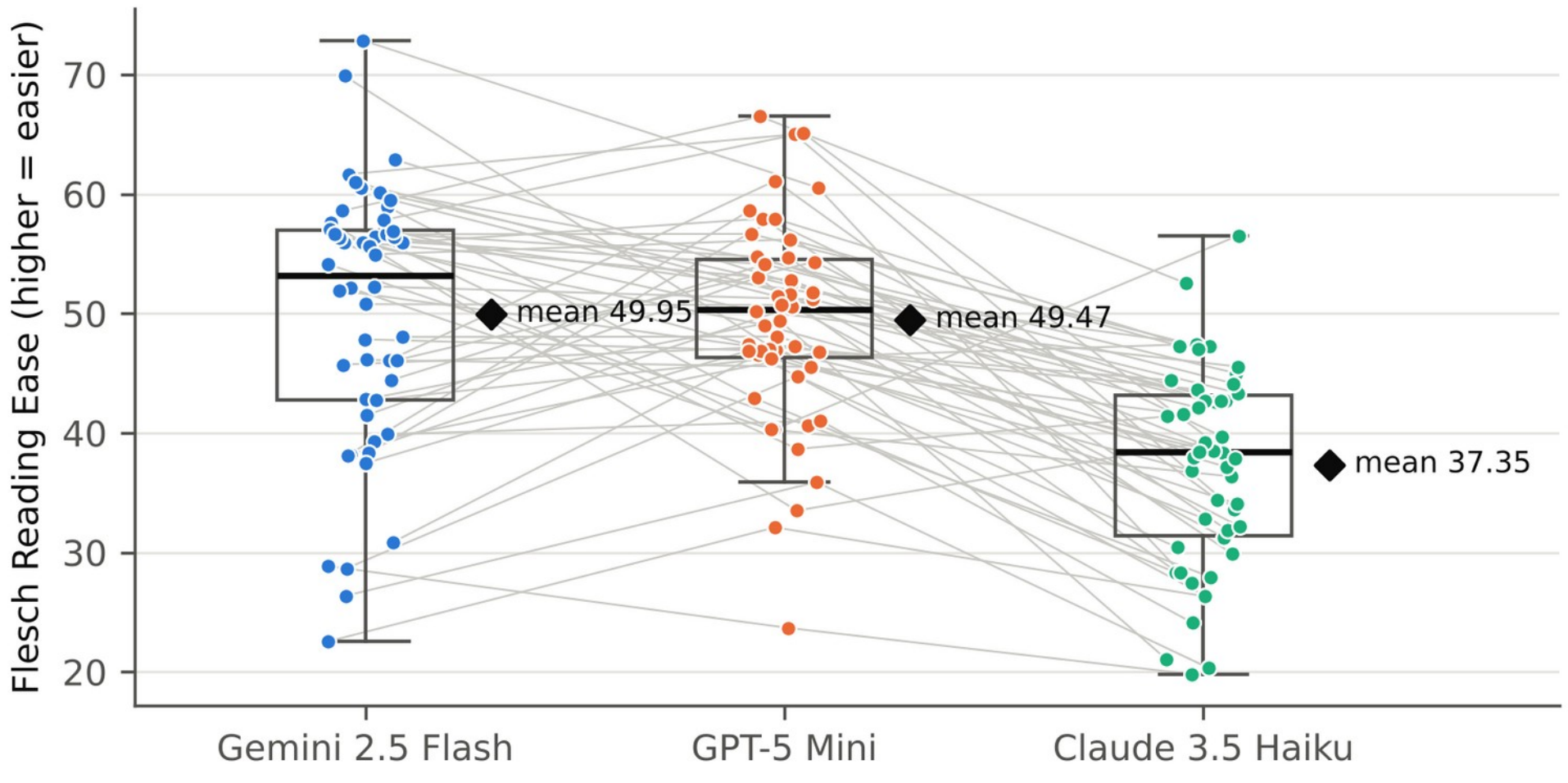

**Fig. 2** Observed Flesch Reading Ease of each description (n = 50 per model). Grey lines join the three descriptions of the same artwork; boxes show the median and interquartile range, whiskers the full range, diamonds the mean.

The models reach their scores in different ways (Fig. 3). Gemini 2.5 Flash usually wrote three sentences of about 19 words each. GPT-5 Mini and Claude 3.5 Haiku almost always wrote two sentences, but GPT-5 Mini's sentences were the longest of the three (28.1 words on average) and its words the shortest (1.52 syllables), whereas Claude 3.5 Haiku used the longest words (1.72 syllables). Relative to GPT-5 Mini, Claude 3.5 Haiku's lower FRE comes entirely from word length: its longer words cost about 16.4 points, partly offset by 4.1 points gained from shorter sentences. Relative to Gemini 2.5 Flash, both components contribute (7.4 points from word length, 5.2 from sentence length). Claude 3.5 Haiku's descriptions were also the shortest (48.5 words, against 55.4 and 57.0). Equal FRE scores can therefore hide different structures: Gemini 2.5 Flash and GPT-5 Mini score alike, but GPT-5 Mini packs the same content into fewer, much longer sentences. Which artworks produced hard text was only weakly consistent across models (Spearman ρ between models 0.29–0.38).

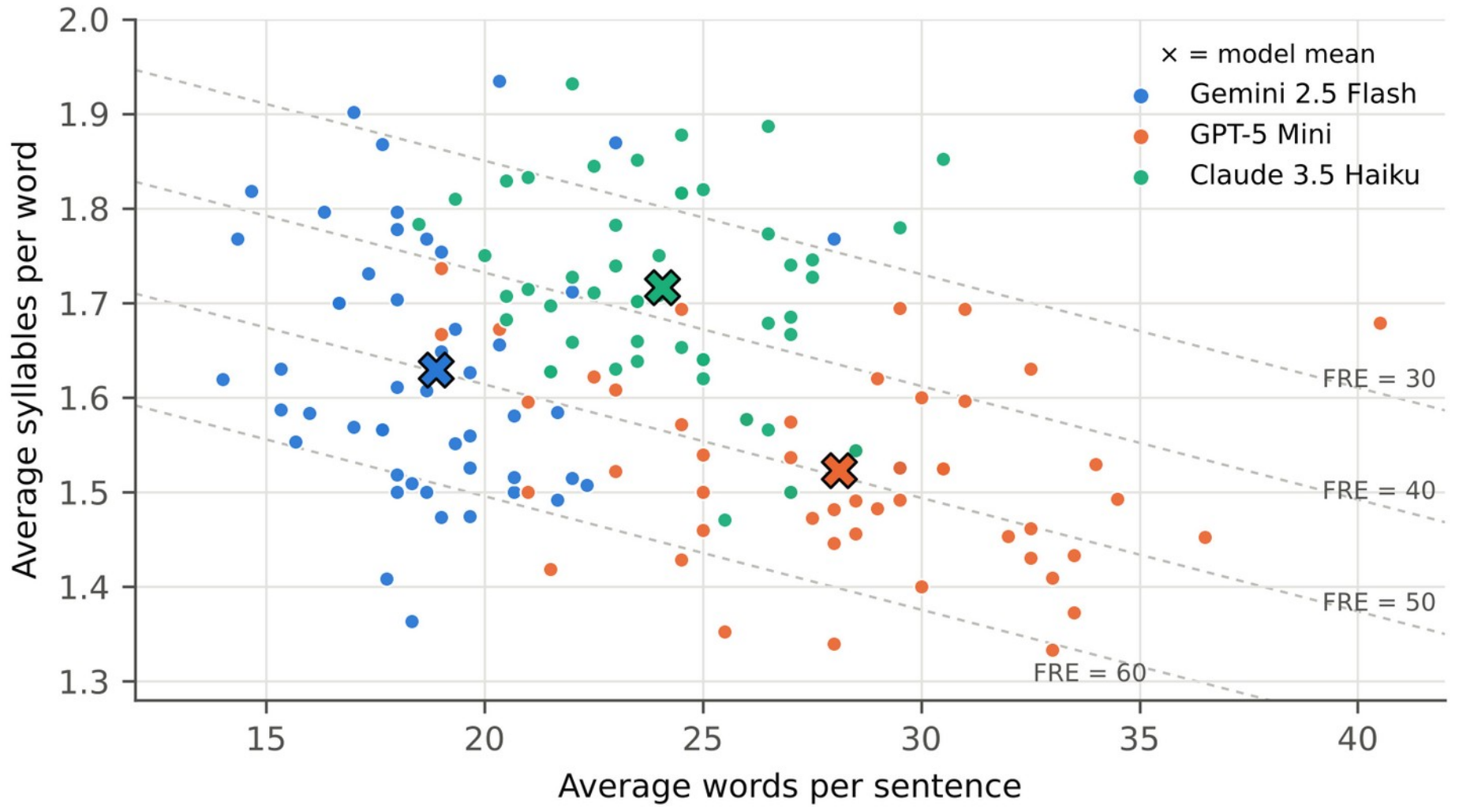

**Fig. 3** The two components of Flesch Reading Ease for every description. Dashed lines join points of equal FRE; crosses mark model means. GPT-5 Mini writes long sentences of short words, Claude 3.5 Haiku shorter sentences of longer words.

### *6.2 Lexical richness*

Lexical diversity and adjective density did not differ between models (Table 1): MATTR 0.912, 0.915 and 0.922 ($F(2, 98) = 1.66$, $p = 0.20$), MTLD 105.0, 113.2 and 110.0 ($F = 1.43$, $p = 0.25$) and adjective density 24.7%, 24.2% and 23.7% ($F = 0.66$, $p = 0.52$) for Gemini 2.5 Flash, GPT-5 Mini and Claude 3.5 Haiku respectively. Roughly one word in four was an adjective in every model's output. The only pairwise difference in this group was a higher MTLD for GPT-5 Mini than for Gemini 2.5 Flash under the Wilcoxon test ($p_{\text{Holm}} = 0.046$), which the paired t-test did not confirm ($p_{\text{Holm}} = 0.075$).

### *6.3 Content coverage (RQ2)*

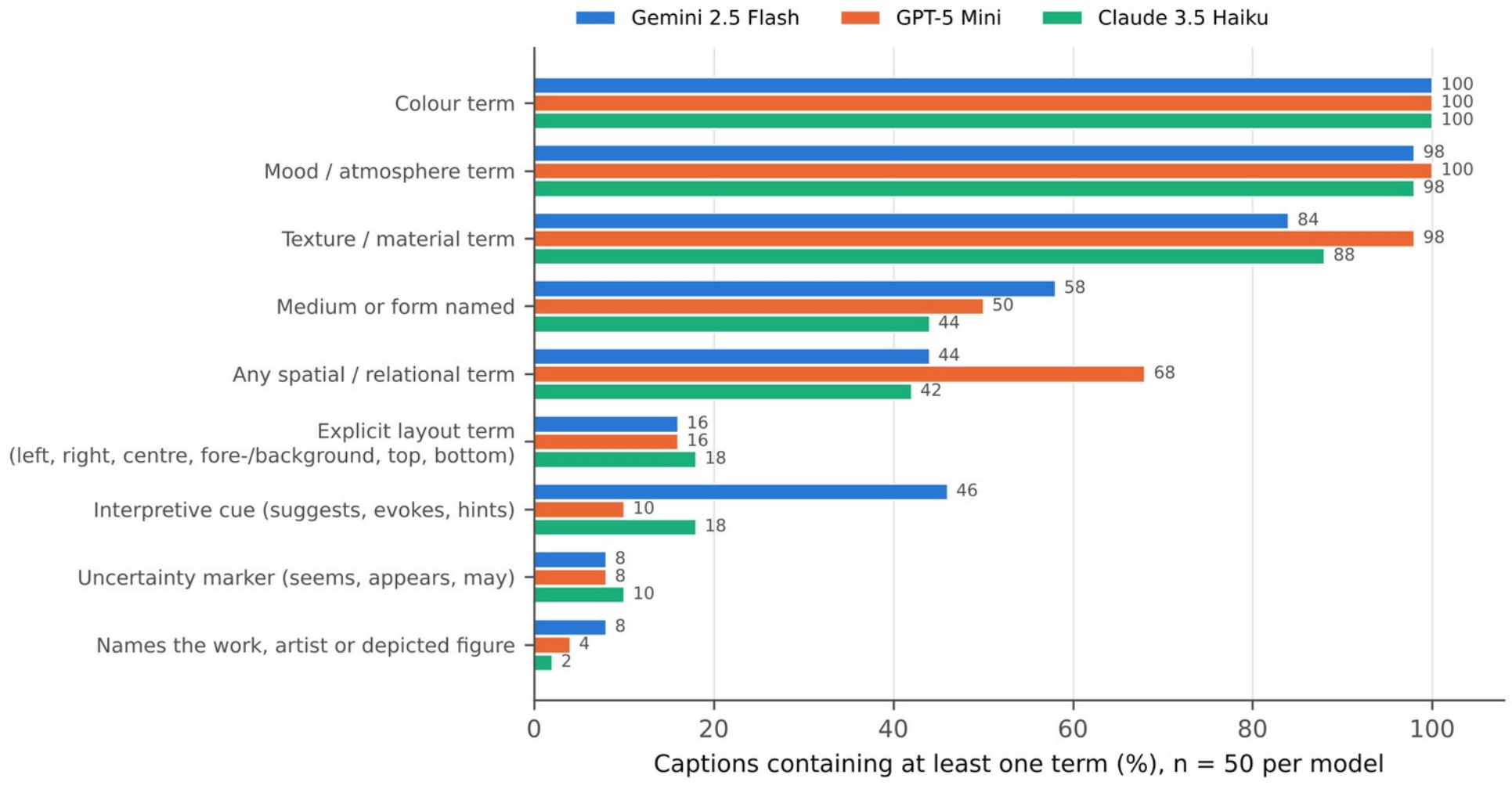

**Fig. 4** Percentage of descriptions containing at least one term from each content dictionary (Appendix B), by model. The prompt requested colour, texture and mood; it did not request layout.

The content the prompt asked for was almost always present (Fig. 4). Every description named colours; 98–100% named a mood or atmosphere; and 84–98% named a texture or material (Cochran's $Q = 6.00$, $p = 0.050$; no pairwise difference survived correction). Between 44% and 58% named the medium or form of the work ($Q = 2.47$, $p = 0.29$).

Spatial information, which the prompt did not request, was much rarer. Only 16% of Gemini 2.5 Flash and GPT-5 Mini descriptions and 18% of Claude 3.5 Haiku descriptions used any explicit layout term such as left, right, centre, foreground or background ($Q = 0.13$, $p = 0.94$). Broader spatial or relational language (e.g. "beneath", "distant", "across") appeared in 44%, 68% and 42% of descriptions respectively ($Q = 10.13$, $p = 0.006$); GPT-5 Mini used it more often than Gemini 2.5 Flash ($p_{\text{Holm}} = 0.045$) and Claude 3.5 Haiku ($p_{\text{Holm}} = 0.013$). A GPT-5 Mini description of *The Garden of Earthly Delights*, for instance, places "a pale-green Eden on the left" and "a central carnival", whereas most descriptions convey colour and mood without saying where anything is (Table 4).

Descriptions rarely signalled uncertainty: 8%, 8% and 10% contained an uncertainty marker such as "seems" or "appears" ($Q = 0.18$, $p = 0.91$). Interpretive cues such as "suggests" or "evokes" were more frequent in Gemini 2.5 Flash descriptions (46%) than in GPT-5 Mini (10%) or Claude 3.5 Haiku (18%) descriptions ($Q = 17.87$, $p < 0.001$; both pairwise $p_{\text{Holm}} \leq 0.013$). Interpretive claims can also appear without any such cue: Claude 3.5 Haiku's description of *The Last Supper*, for example, states that the figures gesture "with a sense of profound, unspoken connection" (Table 4).

### *6.4 Recognition of the specific work*

Seven of the 150 descriptions (4.7%) explicitly identified the work or its subject in a way that requires recognition. Four were from Gemini 2.5 Flash (e.g. "The Last Supper unfolds with Christ and his disciples"; "the philosopher Socrates"). Two were from GPT-5 Mini (Adam and God in *The Creation of Adam*; "Eden" in *The Garden of Earthly Delights*) and one was from Claude 3.5 Haiku, which opened with "Hieronymus Bosch's 'The Garden of Earthly Delights'" ($Q = 2.80$, $p = 0.25$). In addition, six Gemini 2.5 Flash descriptions (12%), one Claude 3.5 Haiku description and no GPT-5 Mini description called the work "iconic" or similar ($Q = 10.33$, $p = 0.006$). Table 4 shows all six descriptions of two works. For *The Last Supper*, one description names Christ while the other two describe the same scene without identifying anyone. For *The Garden of Earthly Delights*, one opens by naming the artist and title, and only one of the three says which panel is on the left or right.

**Table 4** All three descriptions of two artworks, reproduced exactly. Explicit layout terms are in bold; phrases that identify the work, its artist or depicted figures are underlined. All 150 descriptions are in Online Resource 1.

| Model | Description (as generated) | FRE |
|---|---|---|
| **The Last Supper (Leonardo da Vinci)** | | |
| Gemini 2.5 Flash | "The Last Supper unfolds with Christ and his disciples around a long, white-draped table. Deep reds, blues, and earthy tones color their flowing robes as they react with varied emotions, from shock to questioning. The atmosphere is tense and reflective, set against a dark, somber room, with distant windows revealing a serene, blue-green landscape." | 60.99 |
| GPT-5 Mini | "A long, crowded supper table stretches across a dim room where a calm, robed figure in red sits **center**-stage while companions lean, point, and whisper. Muted earth tones and flashes of crimson and green, a crisp white tablecloth, scattered plates, and distant blue windows give the scene a tactile, tense atmosphere — a solemn, charged moment as a group reacts to shocking news at their final meal." | 51.61 |
| Claude 3.5 Haiku | "A warm, hushed scene of intimate gathering unfolds around a long white tablecloth, with rich crimson and earthy green robes draping the figures. Soft light filters through distant windows, casting a contemplative mood as figures lean and gesture with a sense of profound, unspoken connection, creating an atmosphere of quiet anticipation and emotional depth." | 32.16 |
| **The Garden of Earthly Delights (Hieronymus Bosch)** | | |
| Gemini 2.5 Flash | "This triptych transports the viewer from a lush, verdant paradise with soft, pale figures and whimsical pink structures, through a vibrant, crowded garden of earthly pleasures. The journey culminates in a dark, fiery, and desolate hellscape where tortured souls and monstrous creatures writhe amidst harsh, jagged forms, conveying a disturbing descent from innocence to eternal suffering." | 28.85 |
| GPT-5 Mini | "This triptych unfolds in three shifting scenes: a pale-green Eden on the **left** with soft, dewy grasses and quiet figures, a central carnival of fleshy pinks, glossy fruits and shimmering blue pools brimming with naked revelers, and a **right** panel plunged into ashy blacks, glowing embers and jagged, metallic tortures. The textures move from velvety meadow to slick fruit skin to charred, abrasive metal, creating a mood that slides from serene innocence through exuberant, sensual folly to claustrophobic, nightmarish dread." | 23.68 |
| Claude 3.5 Haiku | "Hieronymus Bosch's 'The Garden of Earthly Delights' unfolds like a fever dream, where soft green landscapes dissolve into surreal, densely populated scenes of naked figures, bizarre creatures, and impossible architectural forms. The triptych pulses with an unsettling energy—delicate pastel hues and lush blues contrast with dark, chaotic panels that suggest a journey from innocence through wild indulgence to ultimate darkness." | 19.82 |

### *6.5 Time and cost (RQ3)*

**Table 5** Time and cost: recorded values and estimates

| Item | Gemini 2.5 Flash | GPT-5 Mini | Claude 3.5 Haiku |
|---|---|---|---|
| **Recorded** | | | |
| Run window (UTC) | 5 Nov 2025, 02:35–04:52 | 5 Nov 2025, 05:17–05:48 | 6 Nov 2025, 18:22–18:46 |
| Ten-image batches: first to last narration | n/a (irregular batches) | 9.3–14.6 s (4 batches) | 12.5–31.4 s (4 batches) |
| Model access | US$0.00 | US$0.00 | US$0.10 |
| Characters narrated | 17,424 | 17,182 | 15,859 |
| Text-to-speech (ElevenLabs subscription, all models and pre-tests) | US$5.00 | | |
| Total for 150 narrated descriptions | US$5.10 (about US$0.034 each) | | |
| Not recorded | Per-image upload-to-audio latency; Zapier, Dropbox and Google subscriptions; author time | | |
| **Estimated (manual production, assumptions stated)** | | | |
| Writing a 60-word description at US$0.25 per word | US$15.00 per item | | |
| Narration at US$24.75 per finished minute | US$12.38 (30 s) or US$24.75 (60 s) per item | | |
| Total per item / for 150 items | US$27.38 / US$4,106 (30 s); US$39.75 / US$5,963 (60 s) | | |

Recorded times come from the millisecond timestamps embedded in each ElevenLabs audio ID. A batch is a set of narrations separated by less than 120 s. Rows with a single value apply to the whole study. Manual estimates exclude review and are not like for like: manual descriptions are human-written and reviewed, the automated ones were not reviewed.

Table 5 separates recorded from estimated values. Per-image latency from upload to audio was not logged, and the two-minute polling interval of the Dropbox trigger dominates it, so we do not report it. The narration timestamps do show throughput. In the eight batches in which ten images were added together (four each for GPT-5 Mini and Claude 3.5 Haiku), all ten narrations were completed within 9.3 to 31.4 s of one another, which indicates that the triggered runs were processed concurrently. Gemini 2.5 Flash images were added in smaller, irregular batches, so its timestamps do not support the same calculation.

Recorded spend for all 150 narrated descriptions was US$5.10: no charge for Gemini 2.5 Flash or GPT-5 Mini, which were accessed through AI by Zapier at no additional cost during the study; US$0.10 for Claude 3.5 Haiku; and US$5.00 for an ElevenLabs subscription that also covered pre-tests. This is about US$0.034 per narrated description. The three models produced 50,465 characters of narration in total. Subscriptions to Zapier, Dropbox and Google, and the author's time, are excluded. Because two models were free to access during the study, these figures are a snapshot rather than a price that will hold.

For comparison, we estimated the cost of manual production from two assumptions: writing a 60-word description at US$0.25 per word, and professional narration at US$24.75 per finished minute. These give US$27.38 per item for 30 s of audio and US$39.75 for 60 s, or US$4,106 and US$5,963 for 150 items. These are estimates, not quotes, and the comparison is not like for like: manual descriptions are written and reviewed by people, whereas the automated descriptions here were not reviewed.

## 7 Hallucination, correction and safe deployment

The pipeline's safeguards are operational: the Filter step catches empty output, the alerts catch service failures, and the log makes every output traceable. None of these checks whether a description is true. Our data show three ways in which descriptions could mislead a listener who cannot check them against the image.

First, some descriptions draw on recognition. For famous works such identifications are probably correct, but they come from prior knowledge rather than observation, and because the file names contained the titles we cannot tell whether recognition came from the image or the name. On lesser-known or misattributed works, the same behaviour would produce confident errors. Second, descriptions almost never signal uncertainty (no more than 10% per model), although MacLeod et al. [10] found that BLV users over-trust captions phrased with certainty. Third, interpretive claims can be stated as if they were observed (Sect. 6.3).

We therefore recommend that deployments of this kind add four controls, none of which was evaluated here. (1) Prompt for structure: require one sentence on layout (what is where), and ask the model to keep observation separate from interpretation and to hedge any identification of people, places or events. (2) Flag automatically: the dictionaries in Appendix B can be run on every output to route descriptions with no layout term, with named figures, or with low readability (e.g. FRE below 45) to a reviewer. In the present data, the layout flag alone would have routed 125 of the 150 descriptions to review, which shows how much the prompt, rather than the reviewer, should carry. (3) Keep a person in the loop: descriptions should be checked against the image by a sighted reviewer before publication, with deeper editing reserved for flagship or complex works and the review time recorded so that the true cost of the hybrid workflow is known. (4) Take facts from the record: titles, artists, dates and subjects should come from the collection's catalogue rather than from the model, and the model should be given neutral file names.

## 8 Discussion

### *8.1 What the results mean*

The three models differ in readability, but not in the way a single score suggests. Gemini 2.5 Flash and GPT-5 Mini have the same mean FRE, yet GPT-5 Mini writes sentences about 50% longer. For narration this may matter more than FRE implies, because a listener cannot re-read a long sentence. Claude 3.5 Haiku's lower score comes from longer words rather than longer sentences. A collection that wants plain-language narration should therefore set targets for sentence and word length separately rather than for FRE alone.

Lexical richness was the same across models. Model choice affected sentence structure and word length, not variety, and none of the models produced descriptions that were richer in this sense than the others.

Length is a design decision as much as a model property. An et al. [11] found that BLV participants' ratings of model descriptions varied widely in whether they gave enough detail and whether they were concise. The fixed 90-word limit used here settles that question in the same way for every listener and every work, which suits a museum label but not a listener who wants to explore a painting in depth. The layered design proposed in Sect. 8.2 is one way to let each listener choose.

The cost results matter mainly for scale. At about US$0.034 per narrated description in this study, the marginal cost of producing a first draft is small compared with the cost of checking it. For a museum or library, the economic question is therefore no longer whether descriptions can be generated for a whole collection, but how much reviewer time each description needs before publication. That time is best focused on the works and the content types, such as layout and identification, where errors are most likely. This study did not measure review time; doing so would turn the cost comparison in Table 5 into a like-for-like estimate of a hybrid workflow.

The clearest finding is about content. All three models followed the prompt closely, naming colours, textures and moods in nearly every description, and none often placed elements explicitly in the composition unprompted. That mirrors the weakness Doore et al. [1] found in human-written descriptions, and it means that the sensory emphasis of a prompt determines much of what a BLV listener receives. Spatial layout has to be requested explicitly.

### *8.2 Implications for accessible art-description tools*

For designers of accessible interfaces, the results translate into five requirements. First, structure the request around slots rather than a single free-text instruction: what the work is, its layout (what is where, using left and right, foreground and background), its materials and technique, and finally its mood and possible interpretations. An empty slot can then be detected and flagged before the description reaches a listener. Second, write for the ear. Listeners cannot re-read, so sentence length deserves its own limit; the 28-word average sentences of GPT-5 Mini would be long for narration even though their FRE was acceptable. Where plain language is a goal, WCAG Success Criterion 3.1.5 offers a reference point: when text requires reading ability above lower secondary education, a simpler supplementary version should be available [17].

Third, offer layers. Because the information people want depends on context [9] and blind arts patrons engage with art in varied ways [2], a player can present the short narrated description first and let the listener ask for more detail on layout, technique or history. Fourth, make the status of each statement audible: interpretation should be phrased as interpretation, identification of people or events should be hedged unless it comes from the catalogue, and a description should be labelled as machine-generated until a person has reviewed it. Fifth, keep voice and pace consistent across a collection and let listeners control playback speed, so that differences between descriptions reflect content rather than delivery.

### *8.3 Limitations*

The study has eight main limitations. (1) There was no evaluation with BLV listeners and no assessment of accuracy or hallucination; the content measures show what is mentioned, not whether it is true or useful. (2) The content dictionaries are lexical proxies that miss paraphrases and can match words used in other senses. (3)

Each artwork was described once per model with one prompt, so run-to-run variability and sensitivity to prompt wording are unknown. (4) The artworks are famous and their file names contain titles, which favours recognition; results may differ for lesser-known works. (5) The models are those available through AI by Zapier in November 2025, without exposed version identifiers. Claude 3.5 Haiku belongs to an earlier model generation than the other two, so the comparison describes these specific offerings at that time, not the providers in general. (6) FRE is a reading measure applied to very short texts and depends on the syllable counter. (7) Per-item latency was not logged, and the cost figures reflect free access that may not continue. (8) Like most open collections, the set is dominated by canonical European and North American works; descriptions of work by under-represented artists, regions and media may differ, and widening the set matters if the goal is equitable access.

### *8.4 Future work*

The most important next step is a study with BLV participants that compares descriptions of the same works on comprehension (for example, answering questions about where things are), preference, perceived usefulness and trust, including the effect of hedged wording [10]. Expert annotation of factual errors and unsupported claims would quantify hallucination directly. Re-running the pipeline on lesser-known works with neutral file names would separate description from recognition. Prompt variants that require a layout sentence and separate observation from interpretation, and repeated runs to estimate variability, would show how far these problems can be reduced at the prompt level. Testing several voices would show whether narration quality changes how descriptions are received. For deployment, connecting the pipeline to collection-management systems would let an institution point it at a folder or tag and receive descriptions and audio back, and adding translation and bilingual narration would widen access, provided translations are checked by native speakers.

## 9 Conclusion

CANVAS shows that an end-to-end, no-code pipeline can turn an artwork image into a short narrated description for a few cents, with ten images narrated within about half a minute of one another once a batch was triggered. Under an identical prompt, three contemporary multimodal LLMs produced descriptions that differed in readability and sentence structure but not in lexical richness, and all three described colour, texture and mood far more often than layout. For BLV audiences, that gap matters: a description that conveys atmosphere but not composition makes it harder for a listener to picture the work. Pipelines of this kind are best treated as a fast first draft that requires explicit prompting for spatial structure, automatic flagging and human review, and their value to BLV audiences remains to be established with BLV listeners themselves. For the many collections that currently offer no description at all, a reviewed draft produced in this way could be a substantial improvement in access. The remaining work is to make that review routine and to learn from BLV listeners which descriptions actually help them experience a work of art.

## Supplementary information

Online Resource 1 is a ZIP archive containing: captions.csv (all 150 descriptions with artwork, model, group, input file name, audio URL, ElevenLabs IDs and the recorded FRE score; image URLs are omitted because some works are under copyright); caption_metrics.csv (every per-description measure used in this article); results/ (all descriptive and inferential statistics); and scripts/ (the Python code that reproduces every number, table and figure from captions.csv).

## Declarations

**Funding** This research received no specific grant from any funding agency in the public, commercial or not-for-profit sectors; it was self-funded by the author.

**Competing interests** The author has no relevant financial or non-financial interests to disclose.

**Ethics approval and consent** Not applicable. The study involved no human participants and no personal data.

**Data and code availability** All 150 descriptions, scores, audio links, per-description measures, content dictionaries and analysis code are provided as Online Resource 1, and are openly available at Zenodo (https://doi.org/10.5281/zenodo.23408446) [13].

**Use of generative AI** The study evaluates the output of generative AI models.

## Appendix A The artworks

Each artwork was described once by each of the three models. File names contained the title of the work (e.g. The-Last-Supper.jpg). For Las Meninas, Gemini 2.5 Flash received a JPEG file and the other two models a WebP file; all other inputs were the same file for every model.

**Table A1** The 50 artworks, by group

| Title | Artist | Date | © |
|---|---|---|---|
| **Group 1 Renaissance and Baroque** | | | |
| Las Meninas | Diego Velázquez | 1656 | |
| Judith Slaying Holofernes | Artemisia Gentileschi[1] | c. 1620 | |
| The Garden of Earthly Delights | Hieronymus Bosch | c. 1490–1510 | |
| The Last Supper | Leonardo da Vinci | 1495–1498 | |
| Mona Lisa | Leonardo da Vinci | c. 1503–1519 | |
| The Creation of Adam | Michelangelo | c. 1512 | |
| The Birth of Venus | Sandro Botticelli | c. 1484–1486 | |
| The Night Watch | Rembrandt van Rijn | 1642 | |
| Arnolfini Portrait | Jan van Eyck | 1434 | |
| Girl with a Pearl Earring | Johannes Vermeer | c. 1665 | |
| **Group 2 Impressionism and its successors** | | | |
| Water Lilies | Claude Monet | 1897–1926 (series) | |
| Impression, Sunrise | Claude Monet | 1872 | |
| The Starry Night | Vincent van Gogh | 1889 | |
| The Kiss | Gustav Klimt | 1907–1908 | |
| Wheat Field with Cypresses | Vincent van Gogh | 1889 | |
| A Sunday Afternoon on the Island of La Grande Jatte | Georges Seurat | 1884–1886 | |
| At the Moulin Rouge | Henri de Toulouse-Lautrec | 1892–1895 | |
| The Scream | Edvard Munch | 1893 | |
| Bal du moulin de la Galette | Pierre-Auguste Renoir | 1876 | |
| Where Do We Come From? What Are We? Where Are We Going? | Paul Gauguin | 1897–1898 | |
| **Group 3 Modern and abstract** | | | |
| Broadway Boogie Woogie | Piet Mondrian | 1942–1943 | |
| Nighthawks | Edward Hopper | 1942 | † |
| American Gothic | Grant Wood | 1930 | |
| The Son of Man | René Magritte | 1964 | † |
| Composition VII | Wassily Kandinsky | 1913 | |
| The Persistence of Memory | Salvador Dalí | 1931 | † |
| Black Square | Kazimir Malevich | 1915 | |
| No. 5, 1948 | Jackson Pollock | 1948 | † |
| Les Demoiselles d'Avignon | Pablo Picasso | 1907 | † |
| Guernica | Pablo Picasso | 1937 | † |
| **Group 4 Photography, sculpture, installation and mixed media** | | | |
| One and Three Chairs | Joseph Kosuth | 1965 | † |
| The Dinner Party | Judy Chicago | 1974–1979 | † |
| Tilted Arc | Richard Serra | 1981 | † |
| Drowning Girl | Roy Lichtenstein | 1963 | † |
| Migrant Mother | Dorothea Lange | 1936 | [2] |
| Fountain | Marcel Duchamp | 1917 | † |
| The Great Wave off Kanagawa | Katsushika Hokusai | c. 1831 | |
| A Bar at the Folies-Bergère | Édouard Manet | 1882 | |

| Title | Artist | Date | © |
| --- | --- | --- | --- |
| The Two Fridas | Frida Kahlo | 1939 | † |
| Campbell's Soup Cans | Andy Warhol | 1962 | † |
| **Group 5 Narrative scenes and landscapes** | | | |
| The Oxbow | Thomas Cole | 1836 | |
| Wanderer above the Sea of Fog | Caspar David Friedrich | c. 1818 | |
| The Death of Socrates | Jacques-Louis David | 1787 | |
| The Gleaners | Jean-François Millet | 1857 | |
| Luncheon of the Boating Party | Pierre-Auguste Renoir | 1880–1881 | |
| Christina's World | Andrew Wyeth | 1948 | † |
| Liberty Leading the People | Eugène Delacroix | 1830 | |
| Self-Portrait with Bandaged Ear | Vincent van Gogh | 1889 | |
| The Swing | Jean-Honoré Fragonard | c. 1767 | |
| Whistler's Mother | James McNeill Whistler | 1871 | |

† By an artist who died in 1954 or later or is living; likely still protected by copyright in at least some jurisdictions. These images were used only as model inputs and are not reproduced in this article. [1] The version used was not recorded in the log; the descriptions match the Uffizi version. [2] A US Farm Security Administration photograph, generally regarded as in the public domain.

## Appendix B Content dictionaries

Each description was coded 1 for a dictionary if it contained at least one of the listed terms. The dictionaries are lexical proxies: they record mention, not correctness, and they miss paraphrases. The recognition code (Sect. 5.4) additionally used, for each artwork, the artist's name, non-descriptive titles, and depicted people, places or events (e.g. Christ, Socrates, Eden, the Infanta); the full lists are in Online Resource 1.

**Table B1** Content dictionaries (case-insensitive whole-word matches; ∗ marks a prefix match)

| Dictionary | Terms |
| --- | --- |
| Colour | red(s), crimson, scarlet, vermilion, blue(s), indigo, azure, navy, cobalt, cerulean, green(s), emerald, olive, yellow(s), gold, golden, ochre(s), amber, orange(s), pink(s), rose, purple, violet, lavender, brown(s), umber, sepia, black(s), white(s), ivory, cream, creamy, gray/grey(s), silver(y), teal, turquoise, beige, tan, rust(ed), copper, bronze, monochrome, monochromatic, pastel(s), colo(u)r(s), hue(s), tone(s), palette |
| Texture or material | textur∗, rough, smooth, soft, silky, silk, velvet∗, matte, glossy, gleaming, shimmer∗, grainy, gritty, coarse, brushstroke(s), stroke(s), impasto, thick, layered, crumpled, wrinkled, polished, metallic, rusty, weathered, feathery, fluffy, jagged, sharp, crisp, delicate, tactile, woven, lace, linen, leather, wooden, stone, marble, ceramic, porcelain, glass, steel, canvas |
| Mood or atmosphere | mood, atmosphere, atmospheric, feel(s), feeling, serene, serenity, calm, tranquil, peaceful, tense, tension, melancholy, melancholic, somber/sombre, joyful, joyous, lively, vibrant, eerie, unsettling, haunting, mysterious, mystery, dreamlike, dreamy, ominous, dramatic, solemn, nostalgic, lonely, loneliness, isolation, isolated, anguish, anxiety, anxious, despair, chaos, chaotic, playful, whimsical, intimate, hushed, quiet, contemplative, reflective, sorrow(ful), grief, fear, dread, hope(ful), energy, energetic, emotion∗ |
| Medium or form | painting, painted, paint, oil, fresco, canvas, portrait, self-portrait, triptych, photograph, photo, photography, black-and-white, print, woodblock, woodcut, sculpture, sculptural, installation, readymade, drawing, collage, mural, panel(s), lithograph, comic, pop art, brushstroke(s), brushwork |
| Spatial or relational | left, right, center/centre, centered, central, middle, foreground, background, midground, top, bottom, upper, lower, corner, edge(s), beside, behind, above, below, beneath, underneath, in front of, front, distance, distant, horizon, surround∗, framed by, flank∗, overhead, far, near, nearby, across, along |
| Explicit layout (subset) | left, right, center/centre, centered, middle, foreground, background, midground, top, bottom, upper, lower, corner |
| Uncertainty marker | may, might, perhaps, possibly, appear(s), appearing, seem(s), seemingly, likely, apparently, as if |
| Interpretive cue | suggest∗, evok∗, hint∗, symboli∗, represent∗ |